# MDA-BASED ATL TRANSFORMATION TO GENERATE MVC 2 WEB MODELS

M'hamed Rahmouni[1] and Samir Mbarki[1]

[1]Departement of Computer Science, Faculty of Science, Ibn Tofail University, BP 133, Morocco
md.rahmouni@yahoo.fr
mbarkisamir@hotmail.com

*ABSTRACT*

*Development and maintenance of Web application is still a complex and error-prone process. We need integrated techniques and tool support for automated generation of Web systems and a ready prescription for easy maintenance. The MDA approach proposes an architecture taking into account the development and maintenance of large and complex software. In this paper, we apply MDA approach for generating PSM from UML design to MVC 2Web implementation. That is why we have developed two meta-models handling UML class diagrams and MVC 2 Web applications, then we have to set up transformation rules. These last are expressed in ATL language. To specify the transformation rules (especially CRUD methods) we used a UML profiles. To clearly illustrate the result generated by this transformation, we converted the XMI file generated in an EMF (Eclipse Modeling Framework) model.*

*KEYWORDS*

*MDA, Meta-models, Transformation rules, ATL, MVC 2 Web, UML Profiles*

## 1. INTRODUCTION

Nowadays, it is well recognized that model transformation is at the heart of model driven architecture (MDA) approaches and represents as a consequence one of the most important operation in MDA. The goal of this transformation is to come to a technical solution on a chosen platform from independent business models of any platform [1]. Indeed, the MDA is a new discipline of software engineering that has emerged to ensure the development of productive models [2][3]. MDA brings an important change in the conception of applications taking in account the durability of savoir-faire and of gains of productivity, and taking profits of platforms advantages without suffering of secondary effects[4][5]. Hence, a considerable number of tools supporting the MDA has been developed. On the other hand, many implementations of the MVC 2 pattern, in the field of Web applications, have been carried out. At this level, we cite the following frameworks: Struts [6], Spring MVC [7], php.MVC [8], Zend [9], PureMVC [10]. Among these frameworks, Struts has attained maturity and so has gained the confidence of developers.

This paper presents a tool that transforms a class diagram, which contains CRUD operations, to a model target (especially struts). The result of this transformation is an XMI file that will be subsequently used as a model to generate the source code of a MVC2 Web application. This tool allows to retrieve, delete, update and create the different objects of the information system. The emphasis is placed on the associations between classes. The process of transformation which consists in producing the XMI file is realized using the ATL language (Atlas Transformation Language) [11][12][13].

The remainder of this paper is organized as follows: Section 2 introduces MDA architecture, languages of metamodeling and transformation. Section 3 is devoted to the UML meta-model and MVC 2 Web meta-model. Section 4 presents the specification of CRUD operations by UML profiles. Section 5 describes the process of development, implementation and execution of transformation rules. Section 6 is dedicated to the related work. Finally, section 7 concludes the work and gives hints about future work.

## 2. MODEL-DRIVEN ARCHITECTURE (MDA):

The architecture of MDA is divided into four layers. In the first layer, there are the standard UML (Unified Modeling Language), MOF (Meta-Object Facility) and CWM (Common Warehouse Meta-model). In the next layer, there is the standard XMI (XML Metadata Interchange) which facilitates the exchange of models and their storage. The third layer contains the services that manage events, security and transactions. The final layer provides frameworks adaptable to different types of applications (Finance, Telecom, Transportation, Medical, etc..) [2][3].

The key principle of MDA is the use of models at different phases of application development. Specifically, MDA supports the development of requirements model (CIM), analysis and design (PIM) and code (PSM). The major objective of MDA is to develop perennial models, independent of the technical details of implementation to enable the automatic generation of the entire application code and obtain a significant gain in productivity. In the remainder of this section, we present the main artifacts of the engineering of models, languages expressing the meta-models and the transformation of models: MOF (Meta Object Facility) has been adopted by the OMG in 1997. The MOF specification defines an abstract language and a framework for the specification, construction and management of the generic meta-models. In addition, MOF defines a platform for the implementation of models described by the meta-models [14].

ECORE is a meta-modeling language that is part of EMF (Eclipse Modeling Framework) and is the result of ETP project efforts (Eclipse Tools Project). EMF is a modeling framework and code generation to support the creation of tools and model driven applications. ECORE defines key elements as: EnamedElement, eClassifier, ETypedElement, EPackage, eClass, EDataType, EAttribute, eReference, EOperation.

The model transformation plays a role in the model driven engineering. To this end, several studies have been conducted to define transformation languages effectively and to ensure traceability between the different types of MDA models. ATL (Atlas Transformation Language) is a model transformation language developed in the framework of the ATLAS project [12]. ATL is developed by the team of Jean Bézivin at LINA in Nantes. It is part of the Eclipse M2M (Model-to-Model).

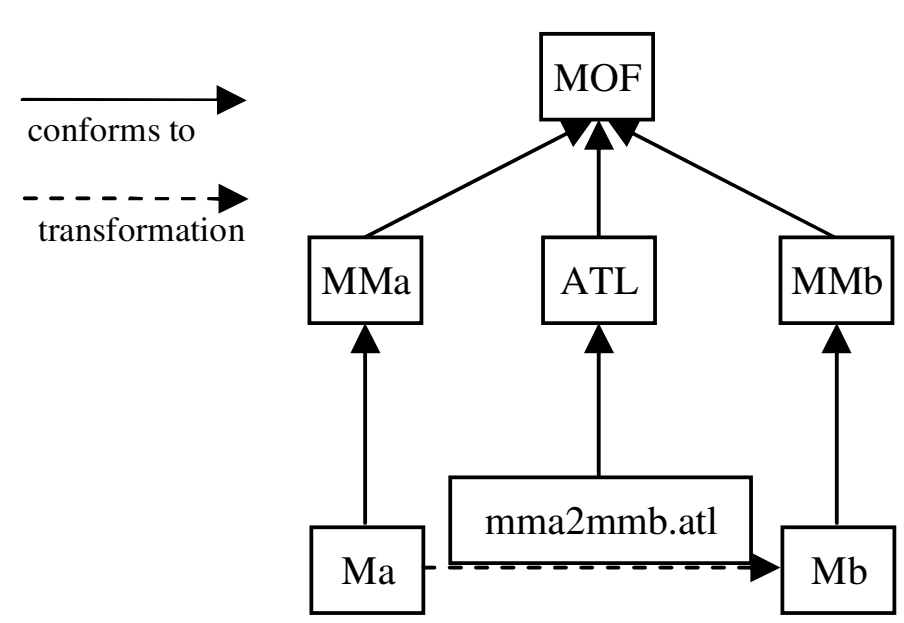

Figure 1. Operational framework of ATL.

## 3. UML AND MVC 2 WEB META-MODELS:

According to the diagram shown in Figure 1, we present in this section the different meta-classes that constitute the PIM and PSM meta-models. Figure 2 illustrates the source meta-model that is a simplified representation of a UML class diagram. UMLPackage corresponds to the concept of UML package, this meta-class is related to the Classifier meta-class. This represents both the concept of UML class and the concept of data type. The Property meta-class expresses the concept of properties of an UML class or references to other classes (uni and bidirectional associations). Figure 3 corresponds to the PSM meta-model. In this meta-model, we show more interest in the tier controller. The ActionMapping meta-class contains the information deployment for a particular Action class. The ActionForm meta-class is a Bean encapsulating the parameters of a form from the view part. The execute() method of Action class performs its processing and then calls the findforward() method on the mapping object . The return value is an object of an ActionForward type. The Action meta-class represents the concept of secondary controller. The Action classes contain the specific processing of the application. Consequently, they must be related to business classes. The complete target meta-model is detailed in [15].

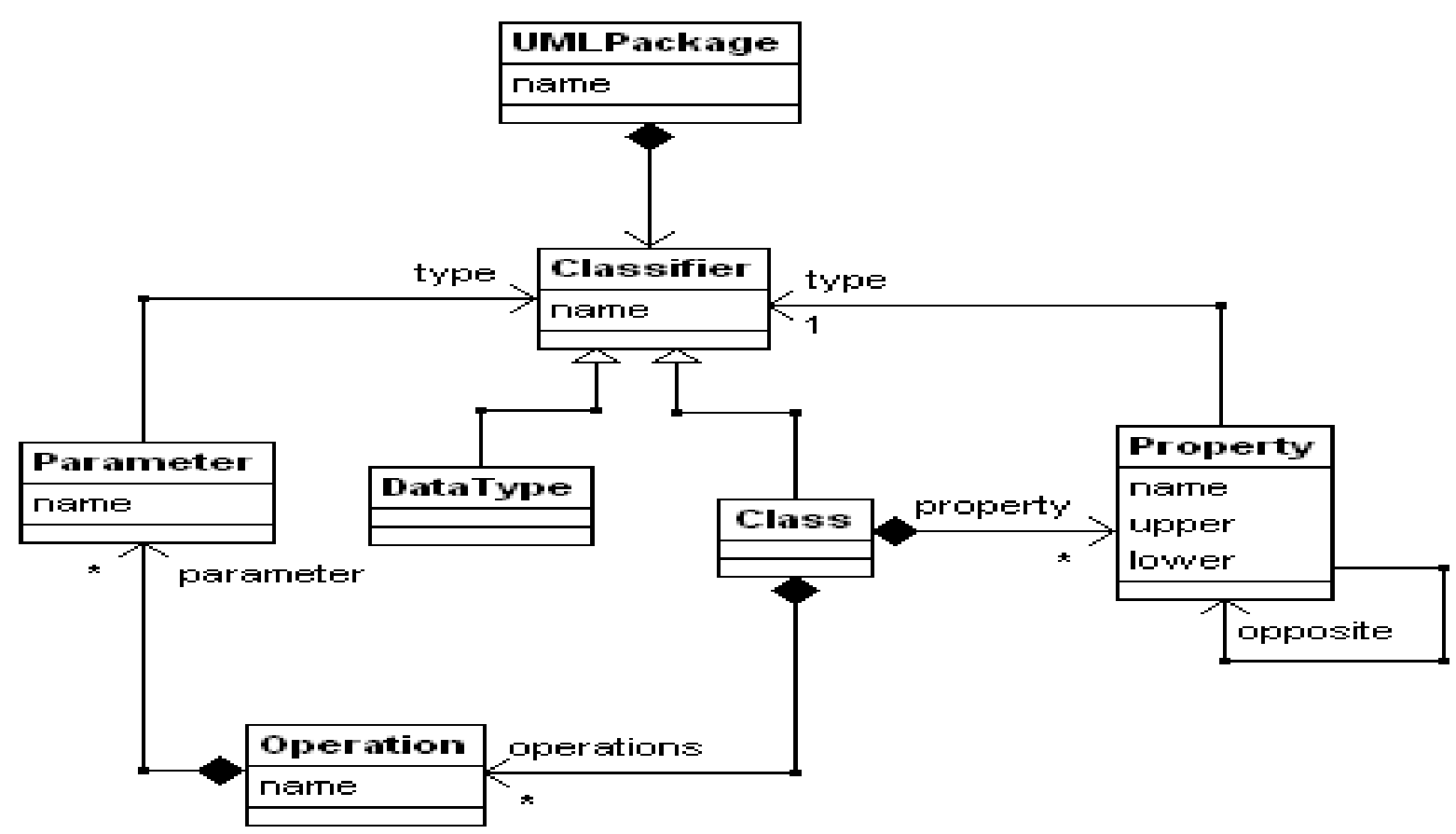

Figure 2.UML meta-model source.

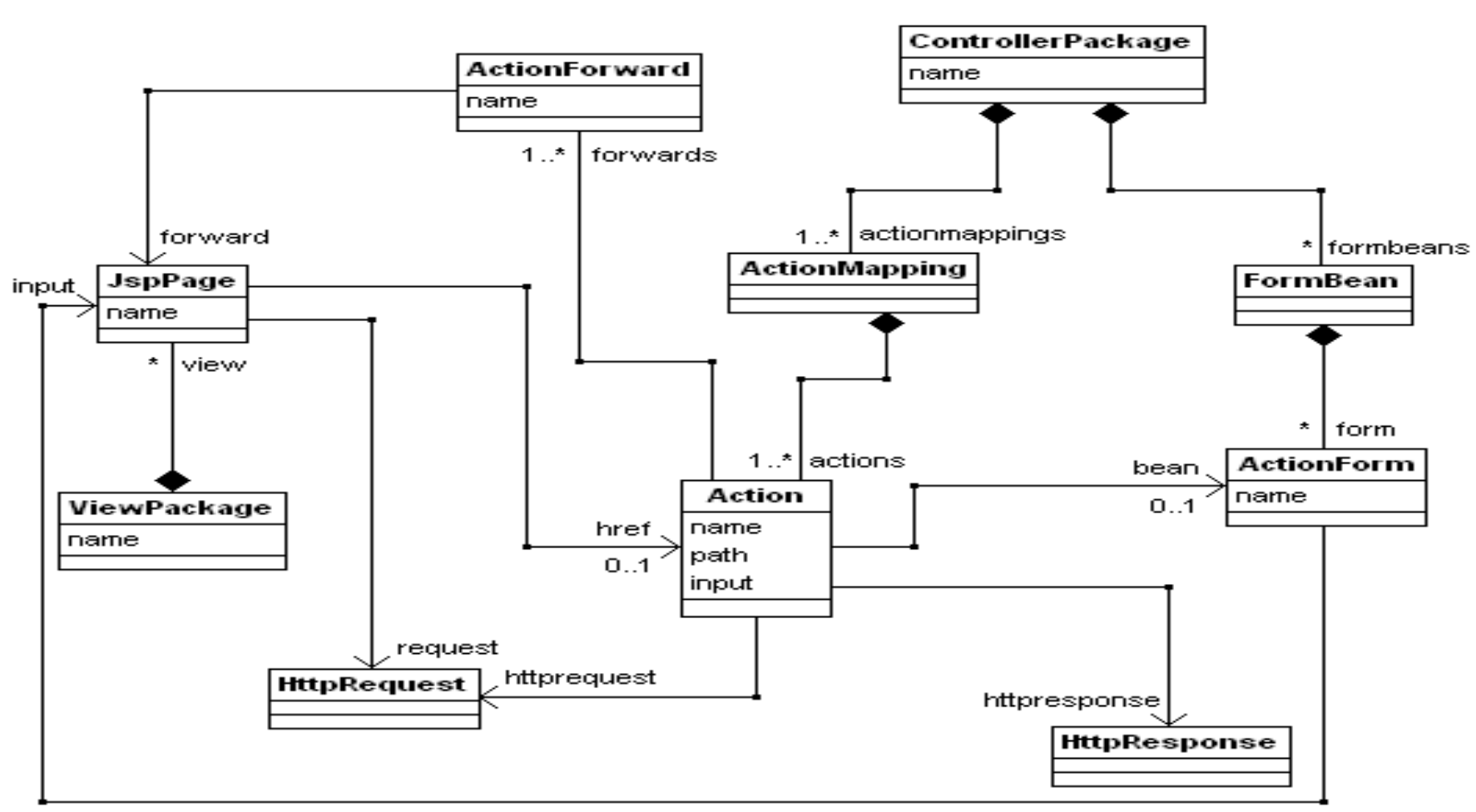

Figure 3.Struts meta-model.

# 4. SPECIFYING CRUD OPERATIONS BY UML PROFILES:

When the UML can not represent a system or application in a practical way, a UML profile can provide extra features to do [16][17]. A UML profile is an adaptation of UML to a particular area [2][18]. It consists of stereotypes, constraints and values marked [19]. A stereotype defines a subclass of one or more elements of the UML meta-model. This subclass has the same structure as its basic elements, but the stereotype specifies constraints and additional properties. In the next section we will specify the operations Create, Retrieve, Update and Delete (CRUD) by UML profiles to predict and specify the outcome of the ATL transformation. The operating algorithm of these operations is based on the principle and architecture of MVC2 Web: Struts [6]. Figure 4 describes this principle.

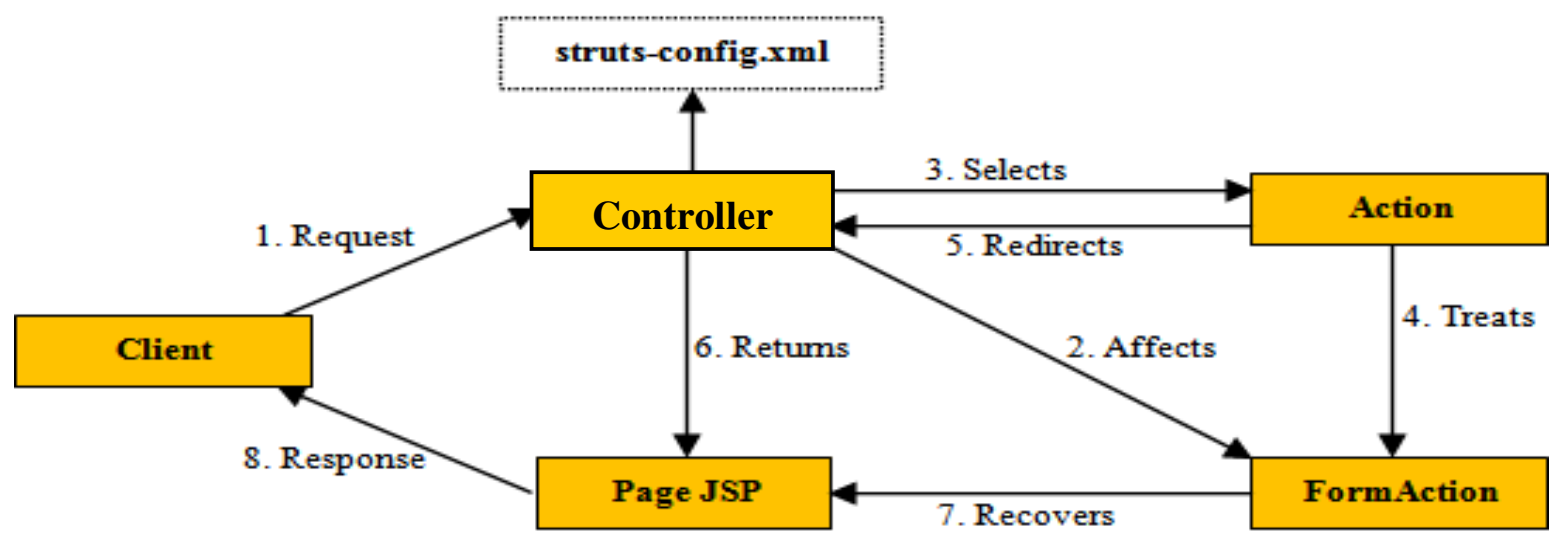

Figure 4. MVC 2 Web Architecture

## 4.1. RetrieveCi Operation

The aim of the Retrieve$C_i$ operation is to display all $C_i$ entities registered in the database already created. This operation is described as follows:

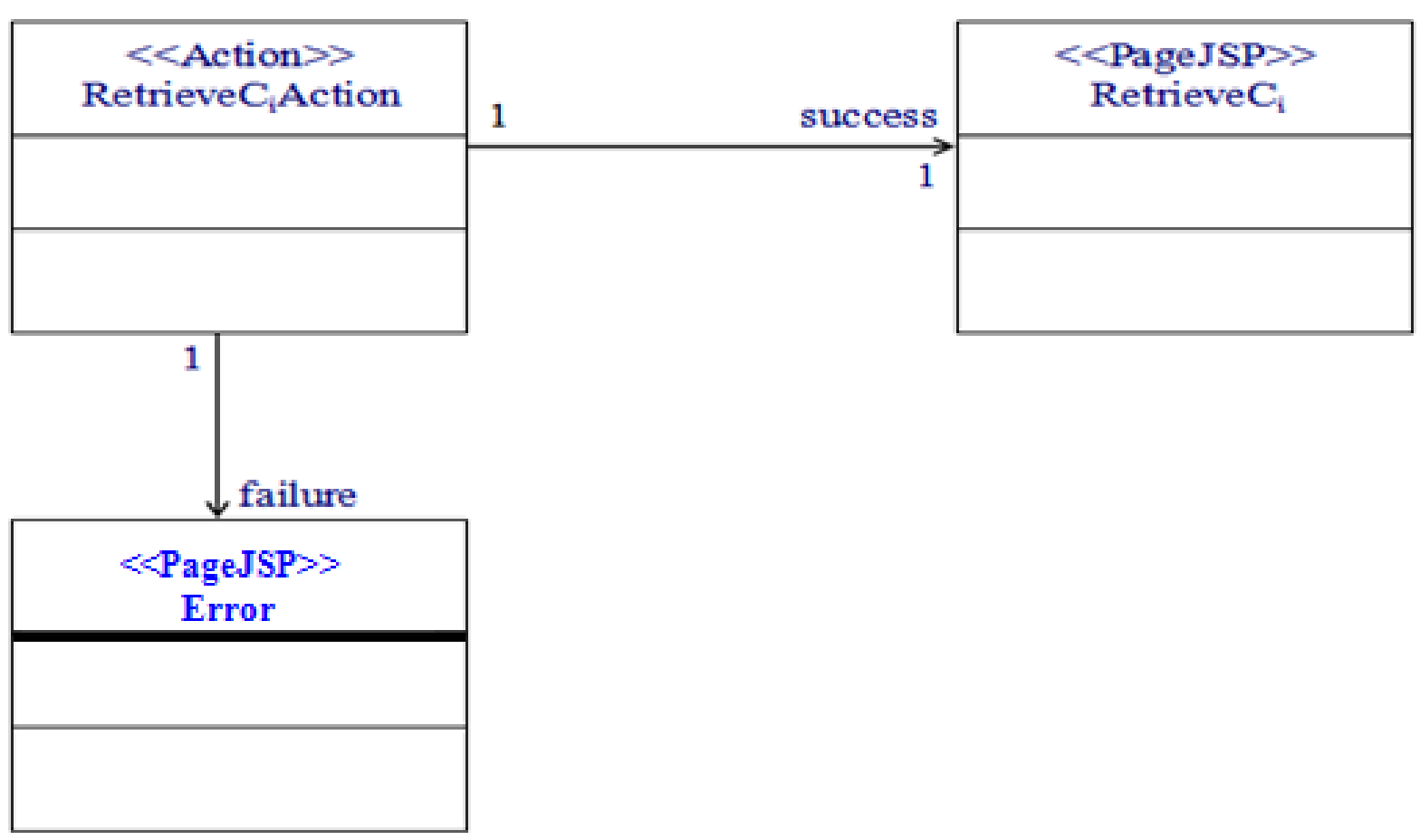

Figure 5. Specifying RetrieveCi operation

## 4.2. The CreateCi Operation

This operation aims at the creation or addition of a new Ci entity in the database. It is described below.

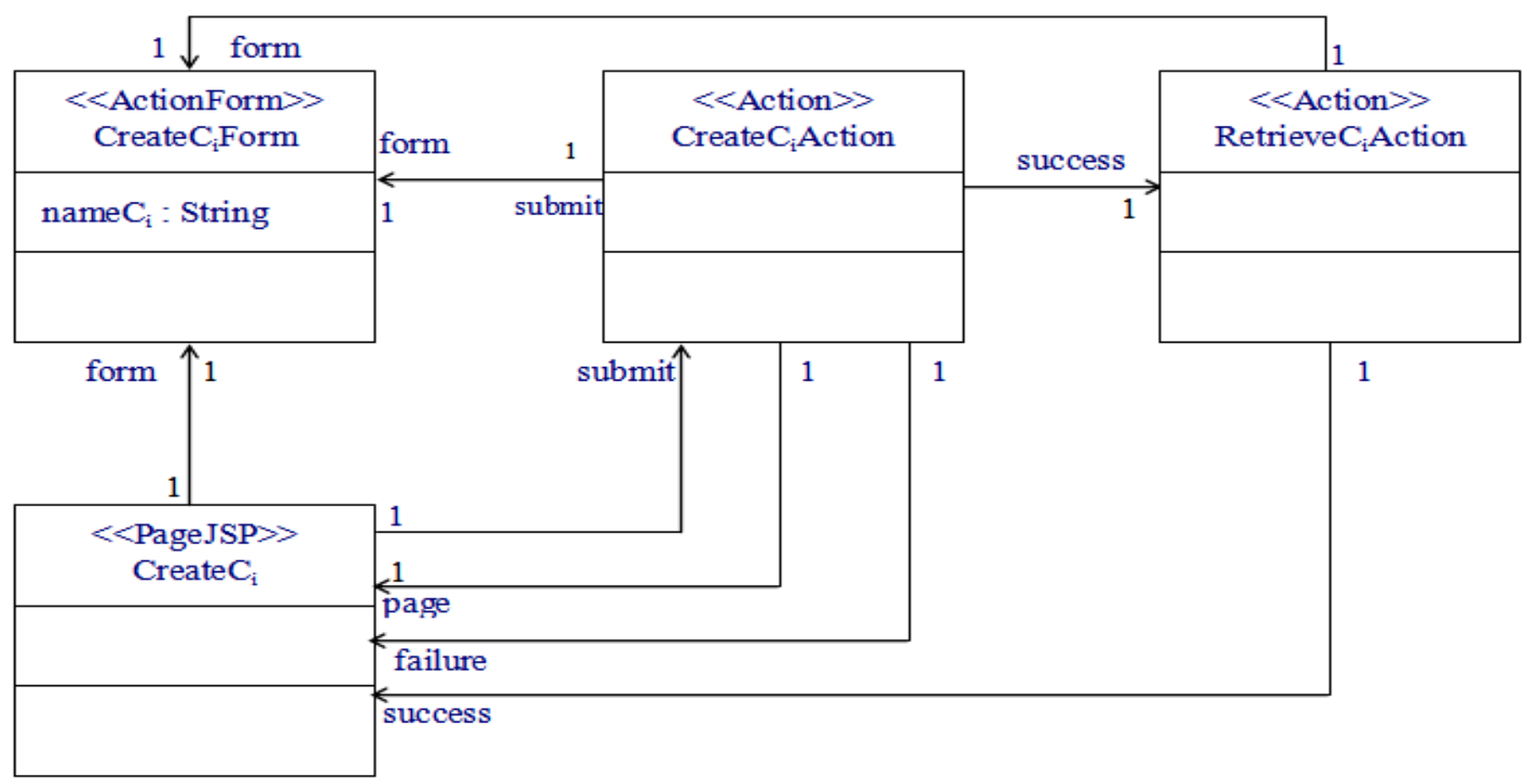

Figure 6. Specifying the CreateCi operation.

## 2.3. UpdateCi Operation

The Ci entity data recorded in the database may be incorrect or false, to correct or change it, we apply the UpdateCi operation detailed as follows by the UML profiles:

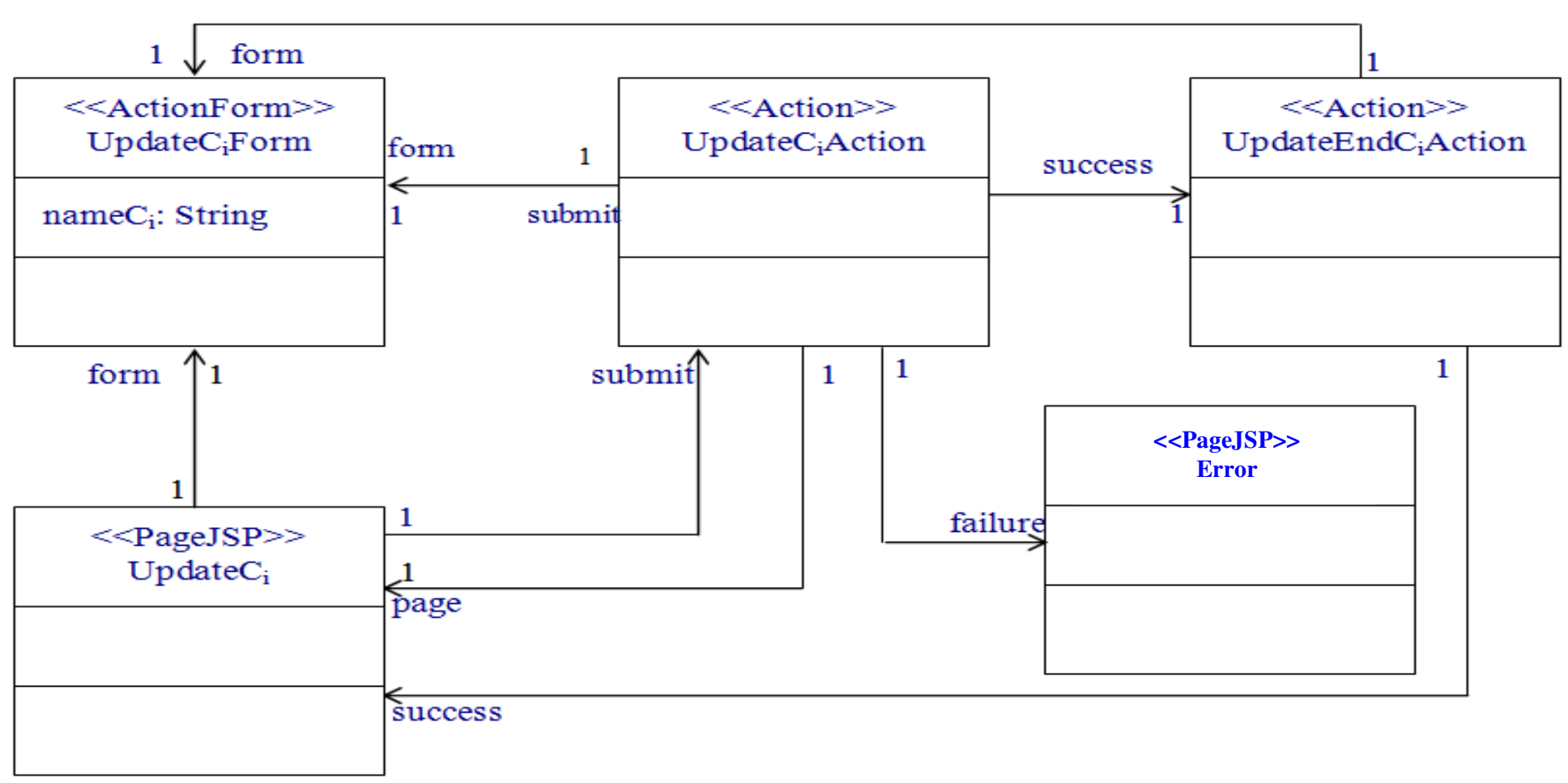

Figure 7. Specifying the CreateCi operation

## 2.4. DeleteCi Operation

The data recorded in the database and which are useless, must be removed. The DeleteCi operation was designed for this role:

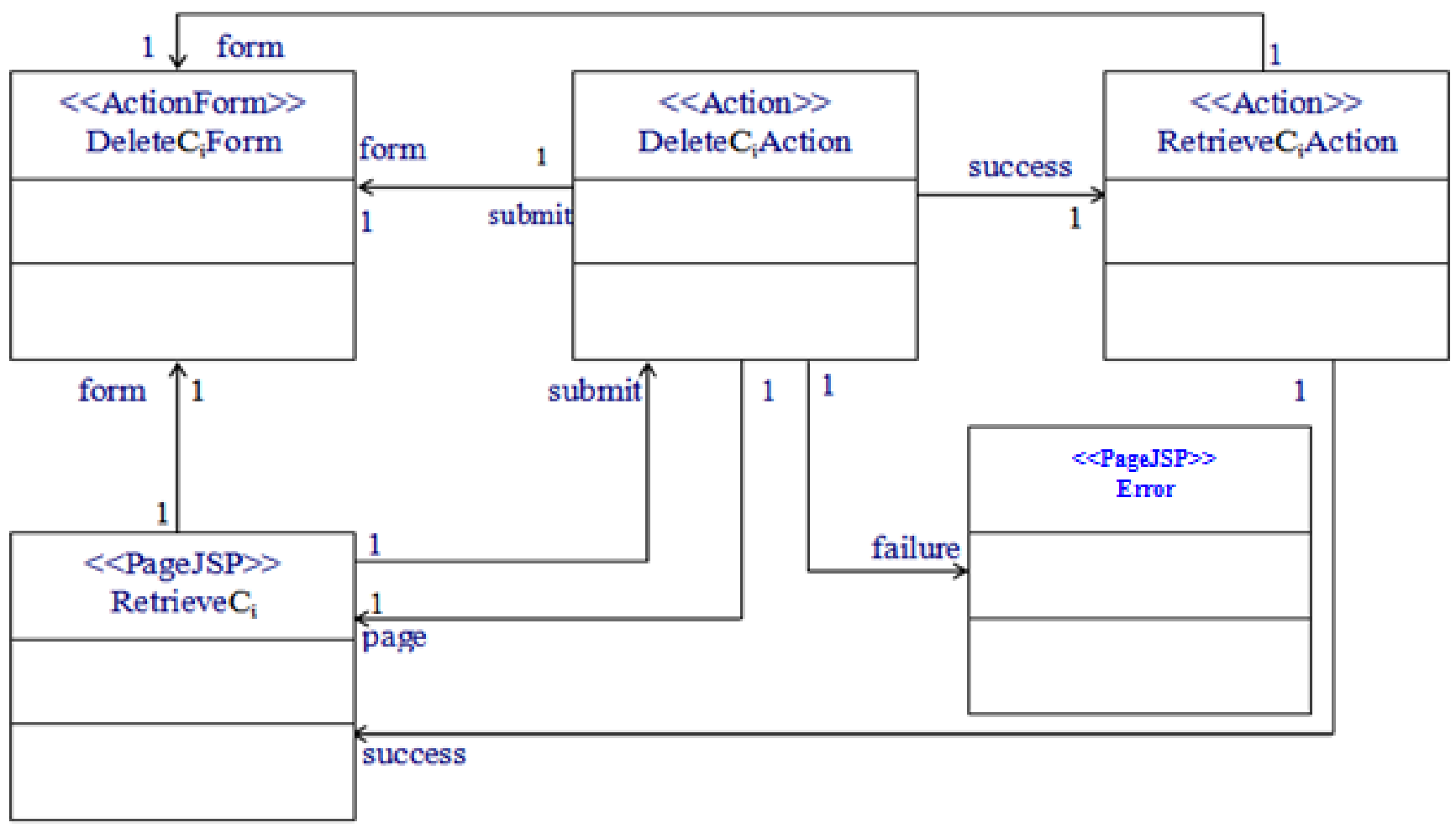

Figure 8. Specifying DeleteCi operation

The specification of other CRUD operations that correspond to Cj and Ck classes will be the same way as that of the Ci class.

## 5. Implementation and Execution of Transformation Rules

In this section, we present the transformation rules allowing to generate a MVC 2 web model from UML class diagram specified by UML profiles and the implementation of these rules then its results after the execution.

In this article we consider only the operations CRUD that Create, Retrieve, Update and Delete, each generates an element of an ActionForm type. The Retrieve operation relating the root class does not generate an ActionForm. The "Create" and "Delete" operations that already performed in [20].

Before proceeding to the execution of the transformation rules, we first created ECORE models corresponding to our two source and target meta-models. In second step, we have implemented the rules of ATL transformation language. The result of XMI file, generated by this transformation, is converted to an EMF model. To validate our transformation rules, we applied more tests. The validation of these rules is also based on the specification of CRUD operations shown above. As an illustration, we consider the UML diagram composed by the classes Ci, Cj and Ck. We get the following XMI model. The target model contains all CRUD operations that: Create, Retrieve, Update and Delete. For the "Retrieve" operation, the Ci root class has no ActionForm. For the "Delete" operation, the input attribute of the "DeleteCiAction" action tag is "/RetrieveCi.jsp" because it is deleting a line from the list of its properties.

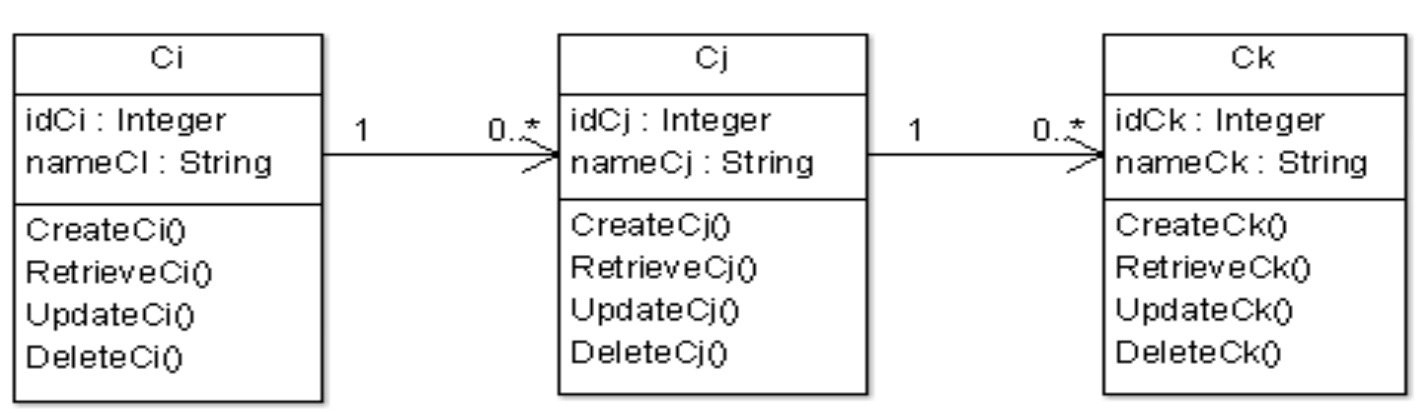

Figure 9. Source Model UML

### 5.1. Rule That Generates the JSP Pages

```
rule P2View{
  from
          a : UML!UML
   to
          vout : STRUTS!ViewPackage(
             name <- a.name,
             view <- Sequence{thisModule.allMethodDefs
                      ->collect(e |  thisModule.resolveTemp(e, 'jsp'))}
                       )
                  }

rule O2JspPage{

         from
         c : UML!Operation
         to
          jsp : STRUTS!JspPage (
                          name <- if c.name='Delete'  then
                                        OclUndefined
                                      Else   c.name+c.class.name+'.jsp'
                                   endif
                                   )
                          }
```

Figure 10. Rule That Generates the JSP Pages

```
<ViewPackage>
   <view name="CreateCi.jsp"/>
    <view name="CreateCj.jsp"/>
    <view name="CreateCk.jsp"/>
   <view name="RetrieveCi.jsp"/>
   <view name="RetrieveCj.jsp"/>
  <view name="RetrieveCk.jsp"/>
    <view name="UpdateCi.jsp"/>
    <view name="UpdateCj.jsp"/>
    <view name="UpdateCk.jsp"/>
</ViewPackage>
```

Figure 11. The Generated result: Package of JSP pages

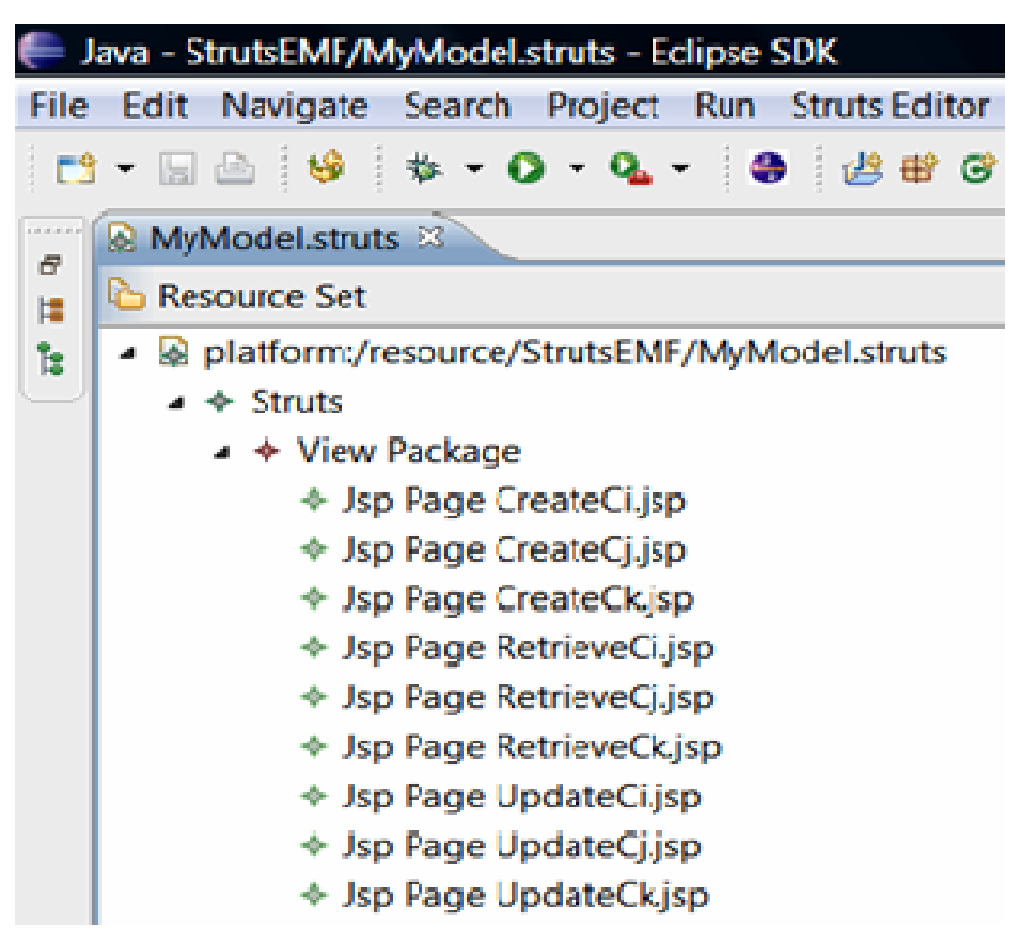

Figure 12. Equivalent result in EMF: Package of JSP Pages

### 5.2. Rule that Generates the Action Classes

```
rule UML2ActionMapping{
    from
        a : UML!UML
    to
        act : STRUTS!ActionMapping(
          name <- 'action'+'-'+'mappings',
          action <- Sequence{thisModule.allMethodDefs
        ->collect(e | thisModule.resolveTemp(e, 'frm')}
          )
      }

rule O2Action{

      from
          c : UML!Operation
      to
           frm : STRUTS!Action(

                path <- '/'+c.name+c.class.name,
                name<- if c.class.opposite.type.name='Void'
                         then if c.name='Retrieve'
                 then OclUndefined
                       else  c.name+c.class.name+'Form'
              endif
          else c.name+c.class.name+'Form'
     endif,
       type <- c.name+c.class.name+'Action',
       input <- if c.class.opposite.type.name='Void'
                  then OclUndefined
              else if c.name='Delete'
                  then   /'+'Retrieve'+c.class.opposite.type.name+'.jsp'
                else    '/'+c.name+c.class.opposite.type.name+'.jsp'
              endif
       endif,
      forward <- Sequence{fr}
       ),

 fr : STRUTS!ActionForward(
         name <- 'Success',
         path <- jsp
               )
}
```

Figure 13. Rule that Generates the Action Classes

```
<actionmappings>
  <action path="/CreateCi" name="CreateCiForm" type="CreateCiAction">
      <forward name="Success" path="/0/@view.0"/>
  </action>
  <action path="/CreateCj" name="CreateCjForm" type="CreateCjAction"  input="/CreateCi.jsp">
      <forward name="Success" path="/0/@view.1"/>
  </action>
  <action path="/CreateCk" name="CreateCkForm" type="CreateCkAction"  input="/CreateCj.jsp">
      <forward name="Success" path="/0/@view.2"/>
  </action>
  <action path="/DeleteCi" name="DeleteCiForm" type="DeleteCiAction">
      <forward name="Success" path="/0/@view.3"/>
  </action>
  <action path="/DeleteCj" name="DeleteCjForm" type="DeleteCjAction"  input="/RetrieveCi.jsp">
      <forward name="Success" path="/0/@view.4"/>
  </action>
  <action path="/DeleteCk" name="DeleteCkForm" type="DeleteCkAction" input="/RetrieveCj.jsp">
      <forward name="Success" path="/0/@view.5"/>
  </action>
  <action path="/RetrieveCi" type="RetrieveCiAction">
      <forward name="Success" path="/0/@view.3"/>
  </action>
  <action path="/RetrieveCj" name="RetrieveCjForm" type="RetrieveCjAction" input="/RetrieveCi.jsp">
      <forward name="Success" path="/0/@view.4"/>
  </action>
  <action path="/RetrieveCk" name="RetrieveCkForm" type="RetrieveCkAction" input="/RetrieveCj.jsp">
     <forward name="Success" path="/0/@view.5"/>
  </action>
  <action path="/UpdateCi" name="UpdateCiForm" type="UpdateCiAction">
     <forward name="Success" path="/0/@view.6"/>
  </action>
  <action path="/UpdateCj" name="UpdateCjForm" type="UpdateCjAction" input="/UpdateCi.jsp">
     <forward name="Success" path="/0/@view.7"/>
   </action>
   <action path="/UpdateCk" name="UpdateCkForm" type="UpdateCkAction" input="/UpdateCj.jsp">
     <forward name="Success" path="/0/@view.8"/>
   </action>
</actionmappings>
```

Figure 14. The Generated result: Package of Action classes

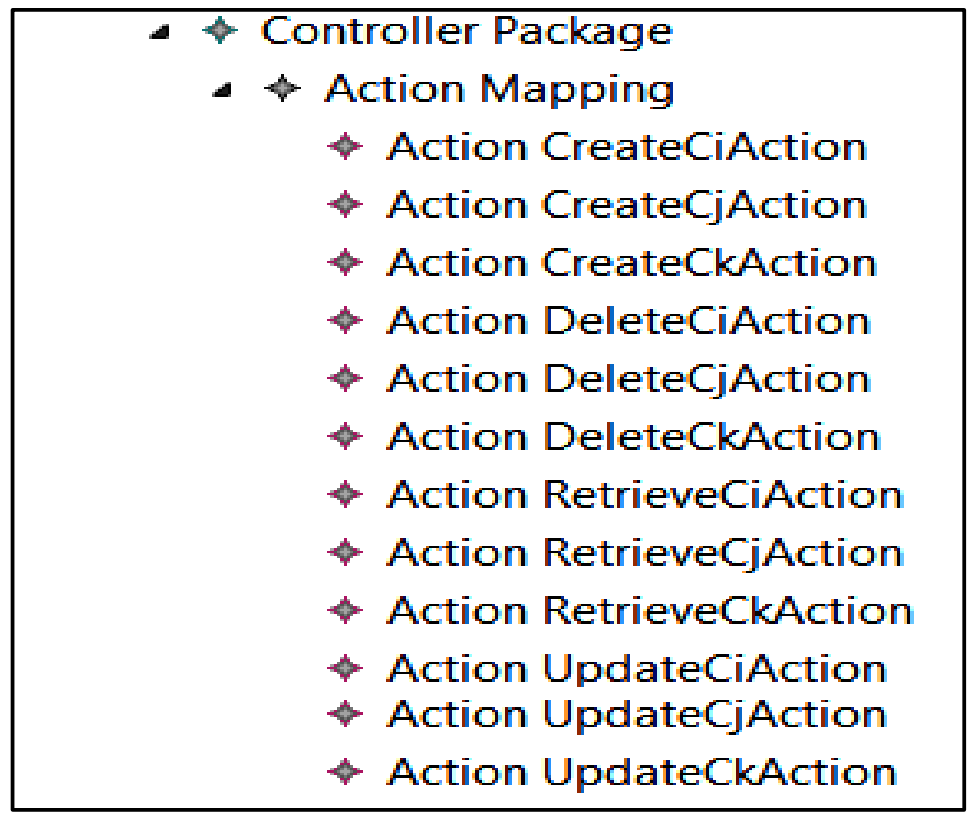

Figure 15. Equivalent result in EMF: Package of Action classes

## 5.3. Rule that Generates the ActionForm Classes

```
rule P2FormBean{
 from
        a : UML!UML
 to

  frmb : STRUTS!FormBean(
        name <- 'form'+'-'+'beans',
        form <- Sequence{thisModule.allMethodDefs ->collect(e | thisModule.resolveTemp(e, 'actf1'))
                       ,thisModule.allMethodDefs ->collect(e | thisModule.resolveTemp(e, 'actf'))}
          )
        }
rule O2ActionForm{

        from
                c : UML!Operation
        to

            actf : STRUTS!ActionForm (
                name <- if  c.class.opposite.type.name='Void'
                        then
                         if c.name='Retrieve'
                            then OclUndefined
                        else   c.name+c.class.name+'Form'
                        endif
                    else c.name+c.class.name+'Form'
                  endif
                 ),

            actf1 : STRUTS!ActionForm (
                name <- if c.name='Create'
                        then   c.name+c.class.name+'End'+'Form'
                     else if  c.name='Update'
                            then  c.name+c.class.name+'End'+'Form'
                         else OclUndefined
                         endif
                    endif
                   )
              }
```

Figure 16. Rule that Generates the ActionForm Classes

```
<formbeans>
    <form name="CreateCiEndForm"/>
    <form name="CreateCjEndForm"/>
    <form name="CreateCkEndForm"/>
    <form name="UpdateCiEndForm"/>
    <form name="UpdateCjEndForm"/>
    <form name="UpdateCkEndForm"/>
    <form name="CreateCiForm"/>
    <form name="CreateCjForm"/>
    <form name="CreateCkForm"/>
    <form name="DeleteCiForm"/>
    <form name="DeleteCjForm"/>
    <form name="DeleteCkForm"/>
    <form name="RetrieveCjForm"/>
    <form name="RetrieveCkForm"/>
    <form name="UpdateCiForm"/>
    <form name="UpdateCjForm"/>
    <form name="UpdateCkForm"/>
</formbeans>
```

Figure 17. The Generated result: Package of ActionForm classes

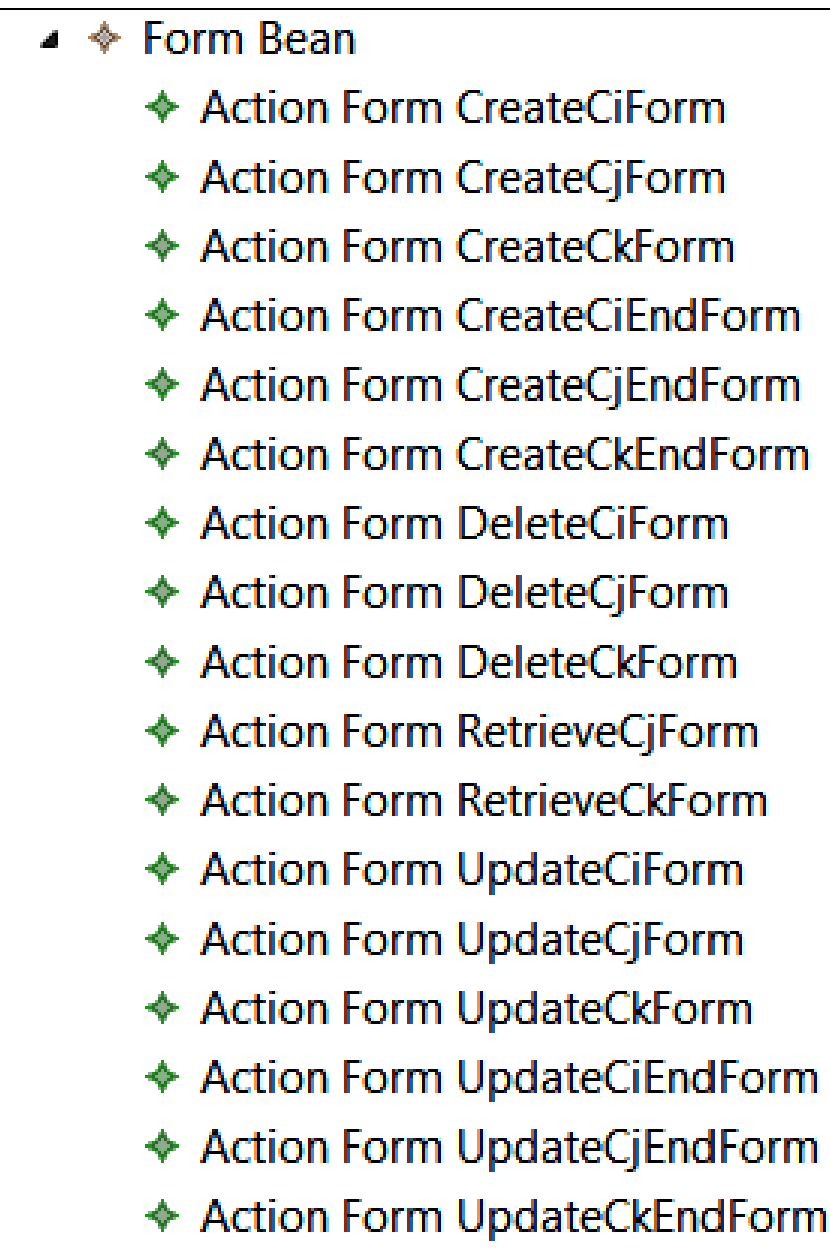

Figure 18. Equivalent result in EMF: Package of ActionForm classes

## 6. RELATED WORK

In the last couple of years we observed a number of proposals for model transformation languages. Many researches on MDE and generation of code have been conducted in recent years. The most relevant are [21] [22] [23] [24] [25] [26] [20] [27] [28].

In [23], the emphasis is placed on generation of code from the requirements. These requirements are expressed as UML use cases. The result of this work is a PSM that can be used to generate a running application. The PSM follows an MVC-based architecture because it generates the code for classes that are called: Controller classes that realize the same use case. The PSM is not developed in a purely MVC 2 logic. We are very interested in this work to generate the code from use case diagram.

A relevant work on the metamodelling was conducted in 2007 [22], the authors developed a meta-model for web requirements. This meta-model takes into consideration concepts like: navigation use case, Web process use case and specific activities such as: browse, search and user transaction. The authors have not developed transformations. The interest of this work can be used as a CIM in order to transform it to a PIM and then to a MVC 2 PSM.

Two other works follow the same logic and have been the subject of two articles [21] [24]. A meta-model for Ajax was defined using the tool AndroMDA. The generation of Ajax code was done and illustrated with a CRUD application that manages persons. The meta-model is very important and we can join it to our meta-models for modeling AJAX user interfaces.

In their paper [25], the authors indicate how to generate JSP pages and JavaBeans using the UWE [29]. The transformation language is ATL [13]. Among the future works cited, the authors consider the integration of Ajax in the engineering process of UWE. UWE has undergone many improvements. Currently, it incorporates models transformation language: ATL and QVT.

The objective of the work of Nasir et al. [27] is to generate the code of a DotNet application" *Student Appointment Management System*". The method used is WebML. The code is generated, applying the MDA approach. Generation is not performed in a DotNet MVC 2 logic.

Another research has focused on aspects of security [28]. A meta-model was developed to integrate the roles of users to access the different pages of the web application. Each page contains navigation rules. Each rule contains a decision (if, elseif, else).

Two proposals are approximates to our work such as [26] and [20]. In [26], the UML meta-model source and the MVC 2 Web meta-model target are built especially for the target from the analysis of MVC 2 Web architecture in a first step and then in a second step the establishment of transformation rules before proceeding to the application by a sample diagram consists of three classes. Each class provided in addition to these properties by a Retrieve operation to display the data and the properties of each class by following an algorithm to display all data. The development of these rules is done via the Java programming leading to an XMI file representing the MVC 2 web model provided only the display operation. By completing this diagram all CRUD operations, and putting a specific algorithm for each operation, before setting new transformation rules between meta-models cited in [26]. Similarly the development of these rules was the Java programming, arriving late to a file in XMI Ecore which is the result of this transformation. The XMI file represents the MVC 2 web model equipped with all the CRUD operations. This is the objective of the work in [20].

Our work adopts the principle of generating the MVC 2 Web models using the ATL transformation language. This generation passes by the specification of the CRUD operations through the use of UML profiles.

## 7. Conclusion and Future Work

Using an MDA approach based on ATL transformation Language we achieve a more automated process for the development of Web applications with a strong focus on architecture modelling. This method can generate the ingredients of web application based on a UML class diagram. The process of generation will provide an opportunity for the user to create, update, delete, and especially to display the different objects it needs. It must be able to display the objects for given class based on information from another object of another class provided that the classes are linked by associations at the class diagram. To do this, we first developed the source meta-model governing UML class diagrams. At the target meta-model, we designed all meta-classes needed to generate a PSM model meets a MVC 2. The transformation rules have been specified by UML profiles and then developed in ATL language. The transformation process can traverse the source class diagram and generating, via these rules, a XMI file containing all actions, forms, forwards and JSP pages that can be used to generate the necessary code to the target application. Further, we plan to generate MVC2 web code from our model, by proposed an automation of our PSM transformation algorithm to code.

## References

[1] L. MENET, (2010) *Formalisation d'une approche d'Ingénierie Dirigée par les Modèles appliquée au domaine de la Gestion des Données de Référence*. PhD thesis, UNIVERSITE DE PARIS VIII, Laboratoire d'Informatique Avancée de Saint-Denis (LIASD), école doctorale: Cognition Langage Interaction (CLI).

[2] X.Blanc, (2005) *MDA en action: Ingénierie logicielle guidée par les modèles*, Eyrolles.

[3] S.J.Mellor, K.Scott, A.Uhl, and D.Weise, (2004) *MDA Distilled: Principales of Model-Driven Architecture*, Addison-Wesley.

[4] A.Atif, A. Jilani, M. Usman, and A. Nadeem, (2011) "Comparative Study on DFD to UML Diagrams Transformations", *World of Computer Science and Information Technology Journal* (WCSIT), Vol.1, No.1, pp 10–16.

[5] N. V. Cuong and X. Qafmolla,( 2011) "Model transformation in web engineering and automated model driven development", *International Journal of Modeling and Optimization*, Vol.1, No.1, pp 7–12.

[6] J.Goodwill, (2002) *Mastering Jakarta Struts*, Wiley.

[7] J.Dubois, J-P.Retaillé, and T.Templier, (2006) *Spring par la pratique*, Eyrolles.

[8] The model view controller framework for php web applications. http://www.phpmvc.net/index.php.

[9] Zend framework. http://framework.zend.com/.

[10] Puremvc framework. http://puremvc.org/.

[11] J.Bézivin, G.Dupé, and F.Joualaut, (2003) "First experiments with the ATL", In *2nd OOPSLA Workshop on Generative Technique in the context of Model Driven Architecture*, Anaheim, USA.

[12] F.Jouault, (2006) *Contribution à l'étude des langages de transformation de modèles*. PhD thesis, Université de Nante.

[13] F.Jouault, F.Allilaire, J.Bézivin, and I.Kurtev, (2008) "ATL : A model transformation tool", *Sciences of Computer Programming-Elseiver*, Vol. 72, No. 1-2, pp 31–39.

[14] Omg/mof Meta Object Facility (mof) specification, omg document ad/97-08-14, (September 1997) (http://www.omg.org).

[15] S.Mbarki and M.Erramdanim, (2008) "Towards automatic generation of MVC 2 web applications", *InfoComp -Journal of Computer Sciences*, Vol. 7, No. 4, pp 84–91.

[16] J. Cabot and C. Gómez, (2003) A simple yet useful approach to implementing UML profiles in current case tools", In *Workshop in Software Model Engineering*, San Francisco, USA.

[17] H. Zhu, G. Li, and L. Zhengmm, (2008) "A UML profile for HLA-based simulation system modeling", In *6th IEEE International Conference on Industrial Informatics*, Daejeon, Korea.

[18] D. C. P. Lopes, (2005) "Etude et applications de l'approche MDA pour des plates formes de Services Web", PhD thesis, Université de Nante, Ecole doctorale: sciences et technologies de l'information et des matériaux.

[19] M.Nassar. Analyse, (2005) "Conception par points de vue : le profil VUML", PhD thesis, Institut national polytechnique de Toulouse, école doctorale: Informatique et Télécommunications.

[20] S.Mbarki, M.Rahmouni, and M.Erramdani, (2010) "Transformation ATL pour la génération de modèles web MVC 2", In *African Conference on Research in Computer Science and Applied Mathematics* (CARI'2010), Yamoussoukro, Côte d'Ivoire.

[21] D.Distante, G.Rossi, and G.Canfora, (2007) "Modeling business processes in web applications: An analysis framework. In *22nd Annual ACM Symposium on Applied Computing*, page 1677, Seoul, Korea.

[22] M.J.Escalona and N.Koch, (2007) "Metamodelling the requirements of web systems", *Lecture Notes in Business Information Processing*, Springer, Vol. 1, pp 267–282.

[23] A.Fatolahi, S.S.Somé, and T.C.Lethbridge, (2008) "Towards a semi-automated model-driven method for the generation of web-based applications uses cases", In *4th Model Driven Web Engineering Workshop*, page 31, Toulouse, France.

[24] V. Gharavi, A.Mesbah, and A.V.Deursen, (2008) "Modeling and generating Ajax applications: A model-driven approach", In *7th International Workshop on Web-Oriented Software technologies*, page 38, New York, USA.

[25] A.Krauss, A.Knapp, and N.Koch, (2007) "Model-driven generation of web application in UWE", In *3rd International Workshop on Model-Driven Web Engineering*, CEUR-WS.

[26] S.Mbarki and M.Erramdani, (2009) "Model-driven transformations: From analysis to MVC 2 web model", *International Review on Computers and Software (I.RE.CO.S)*, Vol. 4, No. 5.

[27] M.Nasir and al., (2009) "WebML and .net architecture for developing students appointment management system", *Journal of applied science*, Vol. 9, No. 8, pp 1432–1440.

[28] E.Oberortner, M.Vasko, and S.Dustdar, (2008) "Towards Modeling role-based Pageflow definitions within web applications", In *4th Model Driven Web Engineering Workshop*, page 1, Toulouse, France.

[29] N.Koch, (2006) "Transformations Techniques in the model-driven developpement process of UWE", In *2nd International Workshop on Model-Driven Web Engineering*, page 3, Palo Alto, USA.

## Authors

**M'hamed Rahmouni:** PhD Student, holding a Diploma of Higher Approfondie Studies in Computer Science and Telecommunication from the faculty of science, Ibn Tofail University, Morocco. He participated in various international congresses in MDA (Model Driven Architecture) integrating new technologies XML, EJB, MVC, Web Services, etc.

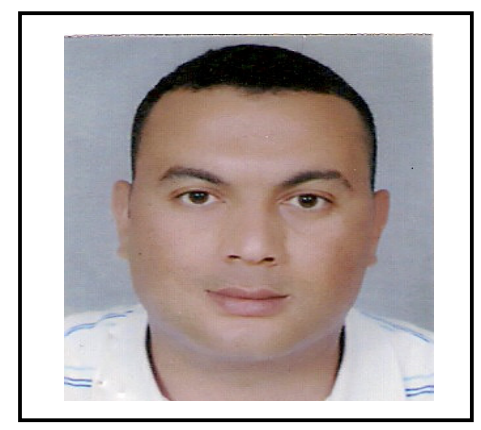

**Samir Mbarki**: Received the B.S. degree in applied mathematics from Mohammed V University, Morocco, 1992, and Doctorat of High Graduate Studies degrees in Computer Sciences from Mohammed V University, Morocco, 1997. In 1995 he joined the faculty of science Ibn Tofail University, Morocco where he is currently a Professor in Department of mathematics and computer science. His research interests include software engineering, model driven architecture, software metrics and software tests. He earned an HDR in computer Science form Ibn Tofail University in 2010.

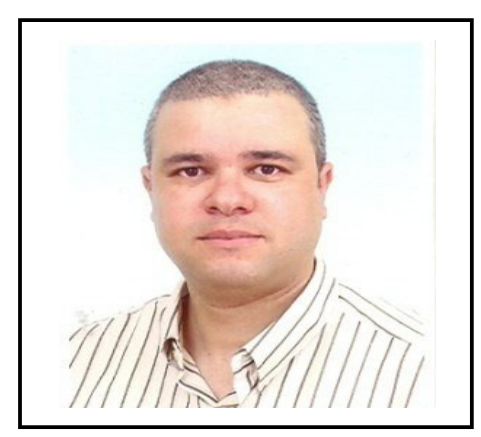